\documentclass[prb,aps,twocolumn,floats,showpacs,amsmath,amssymb]{revtex4}
\usepackage{bm,graphicx}
\newcommand{\be}{\begin{equation}}
\newcommand{\ee}{\end{equation}}
\newcommand{\bea}{\begin{eqnarray}}
\newcommand{\eea}{\end{eqnarray}}
\newcommand{\phrl}[1]{Phys.~Rev.~Lett. {\bf #1}}
\newcommand{\phrb}[1]{Phys.~Rev.~B {\bf #1}}

\newcommand{\bib}{\bibitem}

\begin{document}

\title{Interacting Holstein and extended-Holstein bipolarons}

\author{Monodeep Chakraborty$^1$}
\email[Corresponding Author: ]{monodeep@postech.ac.kr}
\author{Masaki Tezuka$^2$}
\email[]{tezuka@scphys.kyoto-u.ac.jp}
\author{B. I. Min$^1$}
\email[]{bimin@postech.ac.kr}
\affiliation{$^1$Department of Physics, 
Pohang University of Science and Technology, Pohang, 790-784, Korea}
\affiliation{$^2$Department of Physics, Kyoto University,  Kyoto 606-8502, Japan}

%\author{Monodeep Chakraborty}
%\email[E-mail: ]{monodeep@postech.ac.kr}
%\affiliation{Department of Physics, 
%Pohang University of Science and Technology, Pohang, 790-784, Korea}
%\author{B. I. Min  }
%\email[E-mail: ]{bimin@postech.ac.kr}
%\affiliation{Department of Physics, 
%Pohang University of Science and Technology, Pohang, 790-784, Korea}
%\author{Masaki Tezuka}
%\affiliation{Department of Physics, Kyoto University,  Kyoto 606-8502, Japan}
%
\date{\today}


\begin{abstract}
Employing the recently developed self-consistent variational basis generation scheme,
we have investigated the bipolaron-bipolaron interaction within the purview of
Holstein-Hubbard and the extended-Holstein-Hubbard (F2H) model on a
discrete one-dimensional lattice.
The density-matrix renormalization group (DMRG) method has also been used for the Holstein-Hubbard model. 
We have shown that there  exists no bipolaron-bipolaron attraction in the Holstein-Hubbard model. 
In contrast, we have obtained clear-cut bipolaron-bipolaron attraction 
%within the Fr\"{o}hlich paradigm.
in the F2H model.
Composite bipolarons are formed above a critical electron-phonon coupling strength, which can survive the
finite Hubbard $U$ effect.
We have constructed the phase diagram of F2H polarons and bipolarons, and
discussed the phase separation in terms of the formation of composite bipolarons.
\end{abstract}

\maketitle
\section{Introduction}
Bipolarons play an important role in
many areas of material science and biological science.\cite{hohenalder1,Fehske2, Fehske_Trug}
Many body techniques such as the density
matrix renormalization group (DMRG) and the quantum Monte-Carlo (QMC)
%has been employed to investigate Holstein and Fr\"{o}hlich system
have been employed successfully to investigate bipolarons in Holstein and Fr\"{o}hlich systems
at half-filling or quarter-filling.\cite{Fehske2,Masaki,Clay05,Tam,Fehske3,hohenalder2,hohenalder3} The continuum Fr\"{o}hlich bipolarons were also investigated.\cite{dev1,dev2,dev3,dev4}
On the other hand, a proper description of bipolaron-bipolaron ({\it b-b})
interaction in the dilute limit is also very important, as it gives rich information from
entirely different perspective. However, studies on the {\it b-b} interaction 
in the dilute limit are still lacking.
The primary reason is that variational
approaches based on the exact diagonalization (VAED), which have been very successful
for polaron and bipolaron problems,\cite{Trug1,Trug2,Trug3,Trug4,Atis1,Mono1,Mono2} 
can not be applied to even four-electron (two bipolarons) problems. 
Chakraborty {\it et al.}\cite{Mono3} has recently developed 
a new self-consistent scheme of generating variational basis
based on the exact diagonalization (SC-VAED), which makes the {\it b-b}  
interaction problem tractable in the dilute limit. 

In this paper, we have applied the SC-VAED scheme to
Holstein-Hubbard and extended-Holstein-Hubbard (F2H)\cite{Mono2} systems with four and six electrons. 
We have also employed the DMRG method to a four-electron Holstein system
to check the results of the SC-VAED scheme.
Our calculations reveal that the Holstein type of electron-phonon ({\it e-ph}) interaction does not 
yield the attractive interaction between two bipolarons.
In contrast, the {\it e-ph} interaction of extended-Holstein F2 type produces 
the {\it b-b} attractive interaction above a certain {\it e-ph} coupling
to form composite bipolarons.
Hereafter, we name this composite bipolarons as {\it b-composite}.  
The on-site Hubbard electron-electron ({\it e-e}) interaction, $U$,
when it is small,
weakens the inter-bipolaron binding as well
as the intra-bipolaron binding,\cite{Trug2,Mono2} producing a lighter {\it b-composite}. 
Increase in {\it U} leads to the dissociation of the {\it b-composite} 
into two repelling bipolarons, and then into individual polarons. 
However, above a critical {\it e-ph} coupling, 
both the inter-bipolaron and intra-bipolaron bindings survive infinite $U$.
We have constructed the phase diagram of F2H polarons and bipolarons
with respect to {\it e-ph} and {\it e-e} interactions, and discussed the phase separation phenomenon
in terms of the {\it b-composite} formation.

This paper is organized as follows.
In section II, we introduce the Hamiltonian,
which incorporates the {\it e-e} and {\it  e-ph} interactions
within the one-dimensional (1D) Holstein-Hubbard and F2H models.
We also provide computational details here.
In section III, we have examined the bipolaron binding for the four-electron Holstein-Hubbard model,
using the SC-VAED and the DMRG methods.
In section IV, we have made a detailed analysis of the four-electron F2H model 
on the basis of  calculated binding energies and correlation functions in different parameter regimes. 
We have also discussed results of six-electron system within the F2 model.
Then we have constructed the phase diagram of bipolarons with respect to {\it e-ph}
interaction strength $\lambda$ and the on-site Hubbard $U$.
Conclusion follows in Section V.

\section{Hamiltonian and computational details}

                The Fr\"{o}hlich-Hubbard  Hamiltonian on a discrete 1D lattice\cite{Fehske1,Alex1}
is defined as follows:

\bea
H &=&- \sum_{i,\sigma}(t c_{i,\sigma}^{\dag} c_{i+1,\sigma} + \mathrm{h.c.})  
+ \omega \sum_j a_j^{\dag} a_j \nonumber \\
&+& g\omega \sum_{i,j,\sigma}f_{j}(i) n_{i,\sigma} (a_{j}^{\dag} 
+ a_{j}) \nonumber \\
&+& U\sum_{i}n_{i,\uparrow}n_{i,\downarrow}, 
\eea
                              where $c_{i,\sigma}^{\dag}$($c_{i,\sigma}$)
creates (annihilates) an electron of spin $\sigma$  at site $i$
and $a_{j}^{\dag}$ ($a_{j}$) creates (annihilates) a
phonon at site $j$.  We take spin $\frac{1}{2}$ electron ($S_z$=+$\frac{1}{2}$ or $-\frac{1}{2}$). The third term represents the coupling of
an electron at site $i$ with an ion at site $j$ with $g$ being the
dimensionless {\it e-ph} coupling parameter.
$f_{j}(i)$ is the long-range {\it e-ph} interaction, 
the actual form of which is  given by\cite{Fehske1}
\bea
f_{j}(i) = \frac{1}{(|i-j|^{2} +1 )^{\frac{3}{2}}}.
\eea
% If $f_{j}(i)$=$0$ for $i \ne j$, the Fr\"{o}hlich-Hubbard model becomes the Holstein-Hubbard model. 
The {\it e-ph} coupling strength $\lambda$ is defined by\cite{Fehske1,Trug2} 
\bea
\lambda= \frac{\omega g^{2}\sum_{l}f_{l}^{2}(0)}{2t}.
\eea
If $f_{j}(i)$=$0$ for $i \ne j$, the Fr\"{o}hlich-Hubbard model becomes the Holstein-Hubbard model. 
In our study, Fr\"{o}hlich {\it e-ph} interaction is approximated
by the extended-Holstein interaction of F2-type, which corresponds to  the case of $f_{i \pm \frac{1}{2}}(i)$=$1$
and zero otherwise,
as defined by Bon\v{c}a and Trugman\cite{Trug2} and Chakraborty {\it et al.}\cite{Mono2}  F2-model couples an electron
with two nearest-neighbor ions placed in the interstitial.\cite{Trug2,Mono2}
We set the electron hopping $t$=$1$ for numerical calculations,
and so all energy parameters are expressed in units of $t$. 

The variational basis is generated starting with a state of bare electrons and 
adding new states  by repeated application of the
Hamiltonian.\cite{Trug1,Trug3,Atis1,Mono1,Mono2}
 The off-diagonal terms of the Hamiltonian generate different configurations
of phonons and electrons for a specific  up-spin
electron ($\uparrow$) at  position $i$.\cite{Trug1}

All translations of these states 
on the periodic 1D lattice are included.
In the SC-VAED scheme, we first generate a relatively small basis set and obtain the ground state energy and the wave function.
Then the  states with highest probability are identified, and new basis is generated by application
of the Hamiltonian on these chosen highly probable states. Accordingly the size of the basis is increased.  
The ground state energy and the wave function 
is calculated again. This process is continued in a self-consistent way, by increasing the basis size at each cycle 
till the desired accuracy in the ground state energy is obtained.\cite{Mono3}
At each step, the weight of $m$-phonon states for the ground state, $|C_{0}^{m}|^{2}$, as defined
by Fehske {\it et al.}\cite{Fehske1} is calculated. 
The convergence of $|C_{0}^{m}|^{2}$ is checked to ensure that
the basis contains adequate number of phonons required at the given parameter regime.  
 We have considered lattice size up to $L$=$24$ for the four-electron case and $L$=$12$
for six-electron case. In principle, for 4-electrons case, we can go beyond $L$=$24$, probably up to $L$=$32$,
but that would take much more computation time.\cite{CPU}
For the 6-electron case, we can go up to $L$=$18$, but again that would
take much more computation time. 

With obtained ground state energies and wave functions, we discuss the {\it b-b} interaction
in terms of the binding energy ($\Delta$) and the {\it e-e} correlation function ($C_{\uparrow,\sigma}(i-j)$).
$\Delta$ for four- and six-electron systems are  defined by\cite{Trug1,Mono2}
\bea
\Delta_4=E_{4}-2\times E_{2},
\eea
\bea
\Delta_6=E_{6}-E_{4} -E_{2},
\eea
where $E_{2}$, $E_{4}$, and $E_{6}$ are ground state energies of two-electron (one up-spin electron and one down-spin electron), 
four-electron (two up-spin electrons and two down-spin electron) , and six-electron (three up-spin electrons and three down-spin 
electrons) systems, respectively.
$C_{\uparrow,\sigma}(i-j)$ is defined by the probability of finding the other electron(s) 
with respect to one up-spin electron, 
at the position of which the basis is generated.\cite{Trug1,Mono2}
\bea
C_{\uparrow,\sigma}(i-j) = \langle \Psi_{0}\vert n_{i,\uparrow}n_{j,\sigma}\vert \Psi_0\rangle.
\eea
where, $n_{j,\sigma}$ is the number operator with spin $\sigma$ at the
site $j$. In Eq. (6), when $i$=$j$, $\sigma$=$\uparrow$ is not considered
due to Pauli exclusion principle.  $\vert \Psi_{0}\rangle$ is the
ground state wavefunction. 
 Note that, in this work,
we have always considered the same numbers of up-spin and down-spin
electrons and limited the basis states in SC-VAED to ones with total
spin $S$=$0$.

%-------------------------------------------------------------------
\begin{figure}
%\vskip 0.7cm
\includegraphics[scale=0.3]{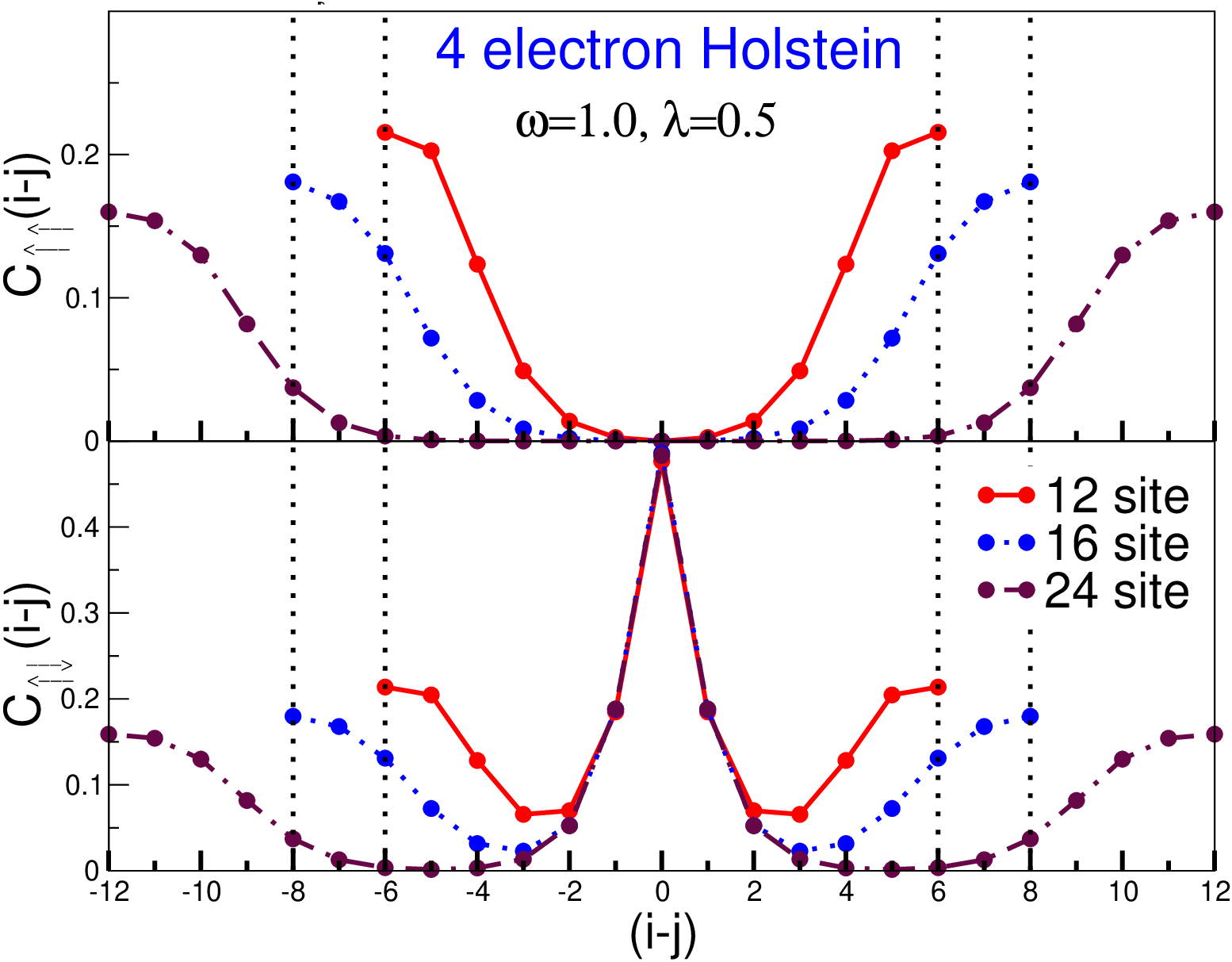} 
\caption{\label{f1} (Color online)
The {\it e-e} correlation function $C_{\uparrow,\sigma}(i-j)$ for the four-electron Holstein lattice ($U=0$). The position $i$ and spin of the electron $\uparrow$ is fixed, since the basis has been generated with respect to it.
$C_{\uparrow^,\sigma}(i-j)$ 's for 12, 16, and 24-site Holstein lattices are compared.  It should be noted, that  $\pm L/2$ for different system sizes are
 the same point.}
\end{figure} 
%-------------------------------------------------------------------

%%%%%%%%%%%%%%%%%%%%%%%%%%%%%%%%%%%%%%%%%%%%%%%%%%%%%%%%%%%%%%
\begin{table}[b]
\caption
{Ground state energies of the four-electron Holstein lattice ($E_4$) for different numbers of lattice sites
obtained using the SC-VAED and the DMRG method. 
$\omega$=$1.0$ and $\lambda$=$0.5$ are fixed.
Converged energy is still higher than twice the ground state energy of the
Holstein bipolaron, $-10.8493$. Energies are in units of $t$.
} 
\label{t1}
\begin{ruledtabular}
\begin{tabular}{c c c}
 Sites & $E_4$ (SC-VAED) & $E_4$ (DMRG)  \\
\hline
12 & $-10.77$ & $-10.782$       \\
16 & $-10.81$ & $-10.797$  \\
24 & $-10.83$ & $-10.828$  \\
60 & $-$      & $-10.846$ \\
\end{tabular}
\end{ruledtabular}
\end{table}
%%%%%%%%%%%%%%%%%%%%%%%%%%%%%%%%%%%%%%%%%%%%%%%%%%%%%%%%%%%%%

%\section{Holstein model}
\section{Holstein-Hubbard model}

     We have calculated ground state energies of four-electron (2 up-spin and 2 down-spin electrons)
Holstein lattice using 
both the SC-VAED and the DMRG method. 
The former was done in the periodic boundary condition,
while the latter\cite{DMRG} was done  in the open-boundary condition.
Results of both methods for different $\omega$ and $\lambda$ 
do not show any signature of binding of the two bipolarons {\it i.e.}, 
two onsite  spin-antiparallel ($S$=$0$)  bipolarons repels each other.
For all parameter regimes, the ground state energies
of the four-electron Holstein system are higher than twice the ground state energy of individual Holstein bipolaron.

Let us look at the obtained ground state energies and
correlation functions for $\omega$=$1.0$, $\lambda$ = $0.5$
as a typical example.
 Twice the ground state energy of a Holstein bipolaron at  this regime is
 $-10.8493$ and
 Table I shows the calculated energies for four electron system at this parameter regime for different lattice sizes. Note that the converged energy is  higher than twice the ground state
energy of the Holstein bipolaron. 
This result is quite expected because an intersite spin-parallel  (S=1) bipolaron, which is formed of two nearest-neighbour electrons with the same spin,
is not favored within the Holstein paradigm.\cite{Trug2,Mono2} The spin-antiparallel bond helps formation of two
bipolarons by binding  up-spin and  down-spin electrons (hence two spin-antiparallel bipolarons are formed from 2 up-spin and
2 down-spin electrons).  However the absence of
 spin-parallel bond  in the Holstein model obliterates the possibility of two bipolarons glueing together.

Figure 1 provides the {\it e-e} correlation function $C_{\uparrow,\sigma}(i-j)$ 
for ground state of the four-electron Holstein system at $\lambda$=$0.5$ and $\omega$=$1.0$.
$C_{\uparrow,\sigma}(i-j)$'s calculated for three different system size (12, 16 and
24-site) show that the two bipolarons 
tend to maintain the maximum separation as far as the system size allows.
 It should be noted, that  $\pm L/2$ for different system sizes are the
same point. Hence,  the sum-rule is satisfied for $-L/2 +1$ to $L/2$ (to avoid
double counting for the end-point).
 
As shown in  Table I, the ground state energy of the four-electron system is lowered
with increasing the system size, which indicates the repulsive nature
of interaction between the two bipolarons. 
 The finite size effect is  evident in this case,
as two bipolarons want to be far apart from each other, but the finite
size of our lattice limits that. Hence, the bigger the lattice size is, the
closer the ground state energy is  to twice the energy of the individual
bipolarons.
The 60-site result by the DMRG method validates this conjecture further (see Table I).
%The DMRG method that can deal with large system size 
%validates this conjecture further with its 60-site result. 
Therefore the formation of  spin-parallel bipolaron is unstable within the Holstein paradigm,\cite{Trug2}
and the scenario remains the same for the four-electron case too. 
The on-site Hubbard $U$ has the same effect on the individual
bipolarons, as reported earlier,\cite{Trug1,Mono2} 
and would not lead to binding between the two  bipolarons.

%-------------------------------------------------------------------
%\begin{figure}[h]
\begin{figure}[t]
\includegraphics[scale=0.33]{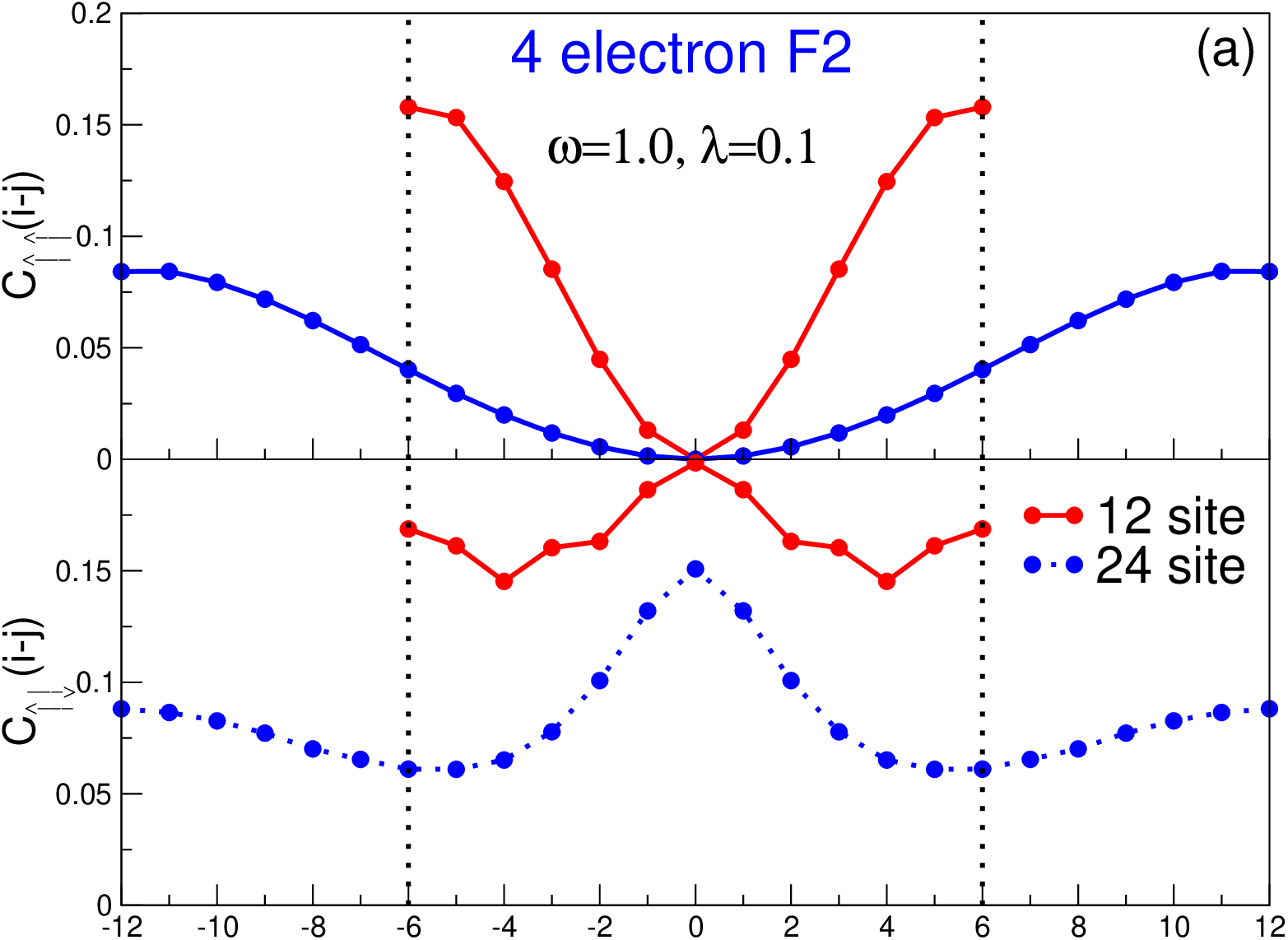} 
\vskip 0.5 cm
\includegraphics[scale=0.325]{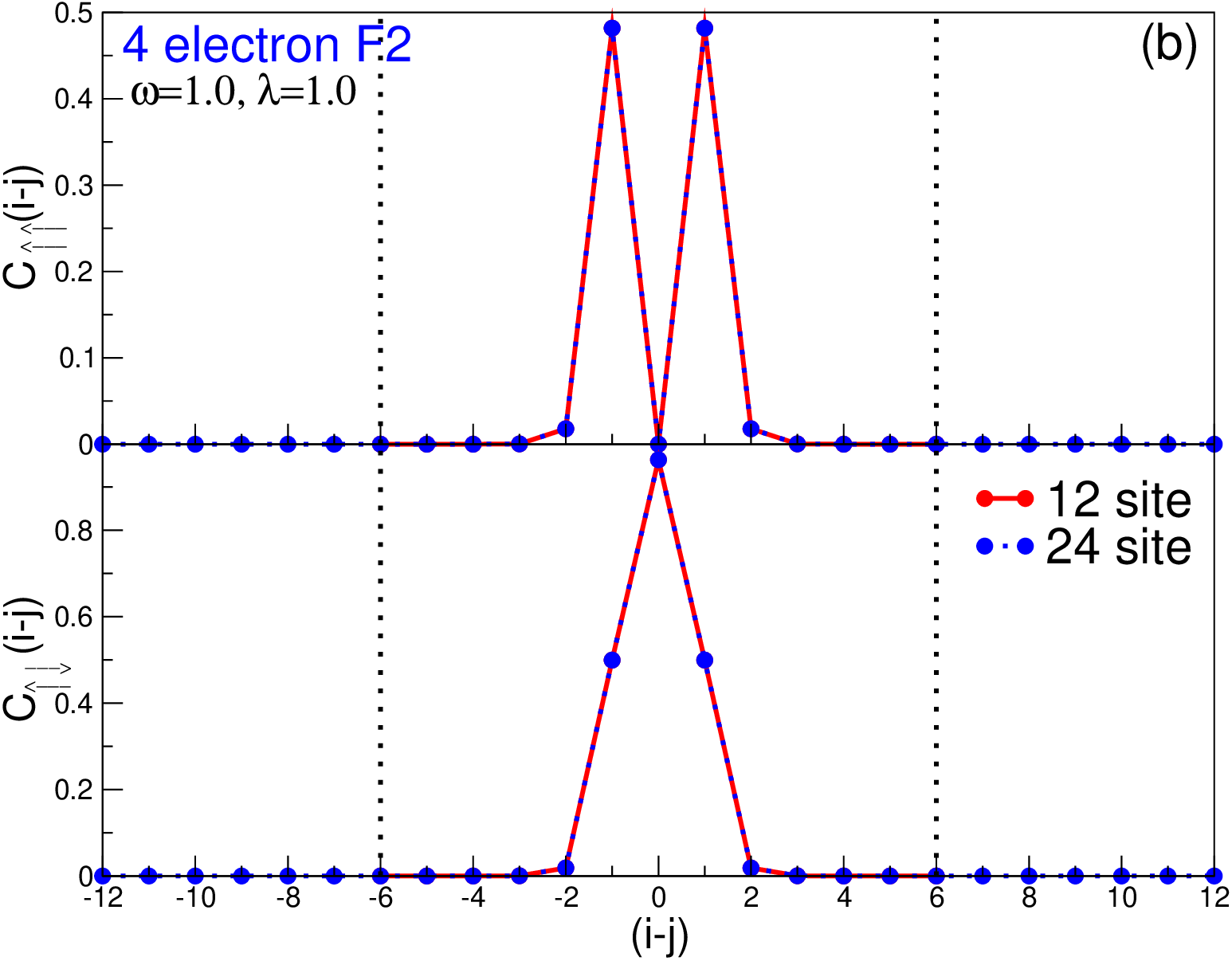} 
\caption{\label{f2} (Color online)
The {\it e-e} correlation function $C_{\uparrow,\sigma}(i-j)$ for the four-electron F2
lattice.  Two different system sizes (12 and 24) are considered. (a) $\omega$=$1.0$, $\lambda$=$0.1$ and (b) $\omega$=$1.0$, $\lambda$=$1.0$.
 It should be noted, that  $\pm L/2$ for different system sizes are the
same point.}
\end{figure}

%\section{F2 model}
\section{Extended-Holstein-Hubbard (F2H) model}

According to earlier calculations,\cite{Trug2,Mono2}
a longer-ranged {\it e-ph} interaction facilitates the binding of the  spin-parallel 
bipolaron above a critical {\it e-ph} coupling strength $\lambda_c^{b}$.
Here we have studied  the four-electron F2H lattice,\cite{Trug2,Mono2}
using the SC-VAED method.  First, we have considered the case of $U$=$0$.
Figure 2(a) shows $C_{\uparrow,\sigma}(i-j)$ for 
$\lambda$=$0.1$ at $\omega$=$1.0$.
In this case, we do not obtain {\it b-b} binding. 
Indeed, as shown in Table II, converged ground state energy of the four-electron 
 F2 lattice for $\lambda$=$0.1$,
$-8.6024$ ($L$=$24$), is higher than twice the ground state energy of the F2 bipolaron, $-8.6215$.
On the contrary, at larger $\lambda$=$1.0$ in Fig. 2(b), $C_{\uparrow,\sigma}(i-j)$ 
clearly demonstrates the formation of a strongly bound {\it b-composite}, 
which is seen to be valid independently of the system size.
In this case, converged ground state energy is $-24.5293$ ($L$=$24$),
which is lower than twice the ground state energy of the F2 bipolaron, $-18.3669$.

Figure 3 shows the binding energy $\Delta_4$ of the four-electron {\it b-composite} 
as a function of $\lambda$.
We see no {\it b-b} binding for small $\lambda$, 
%but
 while, above $\lambda_c^{b-b} \approx 0.3$,
$\Delta_4$ becomes negative, signaling the formation of {\it b-composite}. 
It is notable that, for large $\lambda$, $\Delta_4$ becomes almost
independent, at least up to two decimal places, of the system size (see Table II).
Inset of Fig. 3 shows $\Delta_4$'s for smaller system sizes. 
For small systems of $L$ = $4-8$, {\it b-composite} is apparently formed even for $\lambda < \lambda_c^{b-b}$.  This is
attributed to the finite size effect. With the increase in $\lambda$, the polaronic nature of the individual bipolaron increases, and
due to the small size of the lattice, their wave functions are forced 
to overlap to exhibit a 
spurious binding.
%binding feature. 

%-------------------------------------------------------------------
\begin{figure}
%\vskip 0.5cm
\includegraphics[scale=0.30]{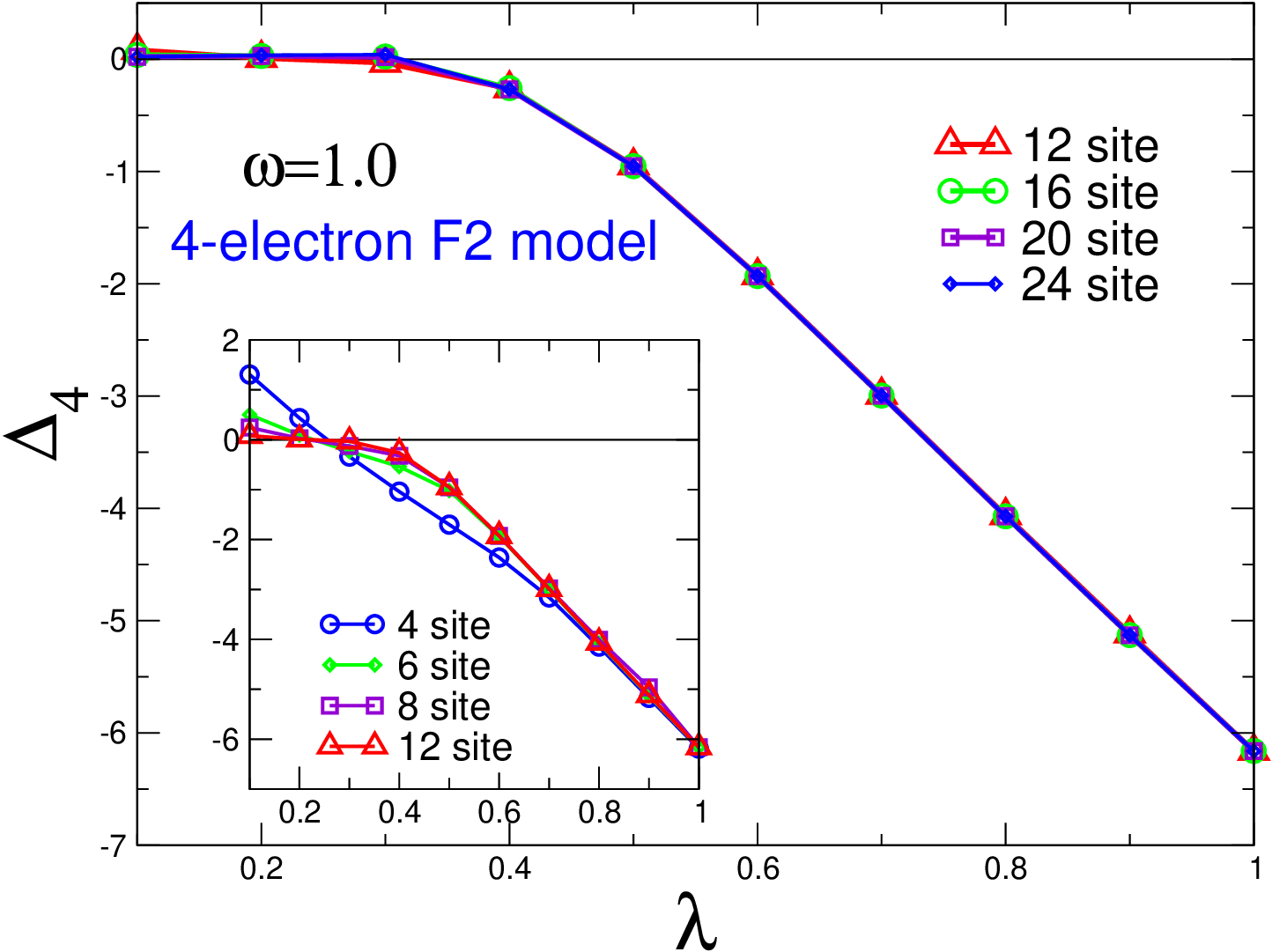} 
\caption{\label{f3} (Color online)
The binding energy ($\Delta_4$) of four-electron F2 {\it b-composite}
as a function of {\it e-ph} coupling strength $\lambda$
for different system sizes. Inset shows $\Delta_4$'s for small system sizes.
$\omega$=$1.0$, $U$=$0$, and energies are in units of $t$.
} 
\end{figure} 
%-------------------------------------------------------------------

Table II presents the ground state energies of four-electron F2 lattices ($E_4$) for different system sizes 
and $\lambda$. Bipolaron (two-electron) ground state energies ($E_2$) at the same parameters are also provided for comparison.
At $\lambda$=$0.1$, as the system size increases, the ground state energy approaches twice the bipolaron energy,
$-8.6215$.  This suggests that the interaction between the two bipolarons  is repulsive. 
Increase in $\lambda$ renders more polaronic nature to these repulsive bipolarons, 
and makes the numerical task more difficult.
Note that, above $\lambda_c^{b-b} \approx 0.3$, the binding energy becomes negative for all system sizes,
and the ground state energy is converged at least up to two decimal places. 
The finite size effect is
 significant in F2 system as well for small  $\lambda$, for which there is
 no {\it b-b} binding for the same reason as mentioned for the Holstein system. 
However, for large
$\lambda$ the finite size effect becomes much smaller because spatial extent of constituent bipolarons and 
that of {\it b-composite} (formed out of coaleasing of two bipolarons) are small. Table II amply
demonstrates this fact.
%-------------------------------------------------------------------

   The energy behaves non-monotonically with $L$ for some values of $\lambda$
in Table II. 
For $\lambda \lesssim 0.3$, the interaction between bipolarons is repulsive and 
the ground state energy of the system with four fermions should converge to twice the
 ground state energy of the system with two fermions (bipolaron); $\Delta$ should converge to zero as $L\to\infty$.

As $\lambda$  increases from zero, our numerical calculation suffers from two
types of difficulties: \\
a) Because of the finite extent (rather large) of the bipolaron, two bipolarons 
can interact with each other along the opposite path as well as along the confronted path,
 when the system size is small under periodic boundary condition. 
  This effect can produce a false binding. \\ 
%as we observe for $L \le 12$ with $\lambda$ = $0.3$.\\
b) When two bipolarons are actually not binding, significantly larger basis size is required 
to get the same accuracy when the system size increases. \\
Therefore, within a realistic computational effort,
it remains very difficult to obtain a universal scaling law of the value of  $\Delta$ with respect to $L$.

On the other hand, for $\lambda \ge 0.4$ where the bipolarons attract each other, while
we observe an
over-estimation of $\Delta$ for very small lattice sizes, the addition of lattice sites
in the periodic boundary condition 
does not significantly increase the numerical complexity. In other words, the effect
we have described above is less 
significant and consequently we can observe that electron-electron correlation
function for $\lambda$=$1.0$  shown in Fig 2(b) shows
a very good convergence between $L$=$12$ and $L$=$24$ and so do their corresponding ground
state energies.
For values of $\lambda$ at which we get a bipolaron composite, the states with phonon
and electrons far away from the composite
centre contribute insignificantly to the ground state, and $\Delta$ (which is negative)
would show a convergence
similar to that of the ground state energies. This situation is the same as in the bipolaron
(bound state of two polarons), where the ground state energy (and hence the $\Delta$) converges up to
seven decimal places with $L$=$37$ 
and remains the same  even for higher $L$.\cite{Trug2,Mono2}

%-------------------------------------------------------------------
\begin{table*}[t]
\caption { 
Ground state energies of four-electron F2 lattices ($E_4$) at different $\lambda$ for different system sizes (L)
($\omega$=$1.0$ is fixed).
Twice the F2 bipolaron energies are also given in the last column.
Energy parameters are in unit of $t$.
}
\begin{ruledtabular}
\begin{tabular}{ c | c  c  c  c  c  c | c }
            & $E_4$ &      &       &      &     &      &$2\times E_2$ \\ \hline
 $\lambda$  & $L$=$4$   &  $L$=$8$ &  $L$=$12$ & $L$=$16$ & $L$=$20$& $L$=$24$ & $L$=$39$ \\  \hline
0.1 & $-$7.3136   & $-$8.3763  & $-$8.5402  & $-$8.5831  & $-$8.5996  & $-$8.6024  &  $-$8.62147    \\
0.2 & $-$8.9795   & $-$9.3858  & $-$9.4028  & $-$9.3840  & $-$9.3835  & $-$9.3776  &  $-$9.41269    \\
0.3 & $-$10.6475  & $-$10.4364 & $-$10.3432 & $-$10.2867 & $-$10.2917 & $-$10.2698 &  $-$10.31085   \\
0.4 & $-$12.3300  & $-$11.6105 & $-$11.5521 & $-$11.5469 & $-$11.5586 & $-$11.5578 &  $-$11.28980  \\
0.5 & $-$14.0364  & $-$13.2887 & $-$13.2831 & $-$13.2876 & $-$13.2882 & $-$13.2902 &  $-$12.33584  \\
0.6 & $-$15.8017  & $-$15.3675 & $-$15.3668 & $-$15.3714 & $-$15.3715 & $-$15.3722 &  $-$13.44075  \\
0.7 & $-$17.7584  & $-$17.5801 & $-$17.5891 & $-$17.5956 & $-$17.5956 & $-$17.5960 &  $-$14.59928  \\
0.8 & $-$19.9535  & $-$19.8156 & $-$19.8709 & $-$19.8791 & $-$19.8792 & $-$19.8793 &  $-$15.80812  \\
0.9 & $-$22.2382  & $-$22.0267 & $-$22.1811 & $-$22.1942 & $-$22.1944 & $-$22.1945 &  $-$17.06493  \\
1.0 & $-$24.5580  & $-$24.5270 & $-$24.5292 & $-$24.5279 & $-$24.5292 & $-$24.5293 &  $-$18.36692  \\  
\end{tabular}
\end{ruledtabular}
\end{table*}
%-------------------------------------------------------------------

Bon\v{c}a and Trugman \cite{Trug2} showed through their analytical and numerical considerations
that the F2 spin-parallel ($S$=$1$) bipolaron is stabilized above $\lambda_c^b$=$0.76$ at $\omega$=$1.0$. 
On the other hand, we have found, that in the above case of four-electron system,
the two bipolarons  binds above $\lambda_c^{b-b}$=$0.3$, which is much lower than $\lambda_c^b$=$0.76$. 
The reason why the {\it b-b} binding takes place at lower $\lambda$
is due to the 
presence of intersite ($S1$)  spin-antiparallel  bond between the two  adjacent on-site  ($S0$) bipolarons which stabilizes the  spin-parallel bond
between them.

We substantiate this behavior with our six-electron F2 lattice calculations.
Figure 4 shows $\Delta_6$'s of the six-electron F2 system as a function of $\lambda$.
%It is seen that {\it b-composite} is  to be formed for 
It is seen that a {\it b-composite} of three bipolarons is to be formed for 
$\lambda \gtrsim 0.3$.
%$\lambda > \sim0.3$.
The numerical precision is not as high as the four-electron case, and especially
lower for the region marked by the ellipse. 
Once there occurs a {\it b-b} binding, the numerical precision is improved,
but at best up to one$-$two decimal places. 
Better numerical precision can be obtained, but that would be too expensive (basis states of the order of $2-3\times 10^7$) 
and time consuming (few weeks to a month). 
In any way, the six-electron result qualitatively validates the fact that the bipolarons
glue into a {\it b-composite} above a certain $\lambda \sim0.3$ (for $\omega$=$1.0$). 

%-------------------------------------------------------------------
%\begin{figure}[h]
\begin{figure}[b]
\vskip 0.6cm
\includegraphics[scale=0.30]{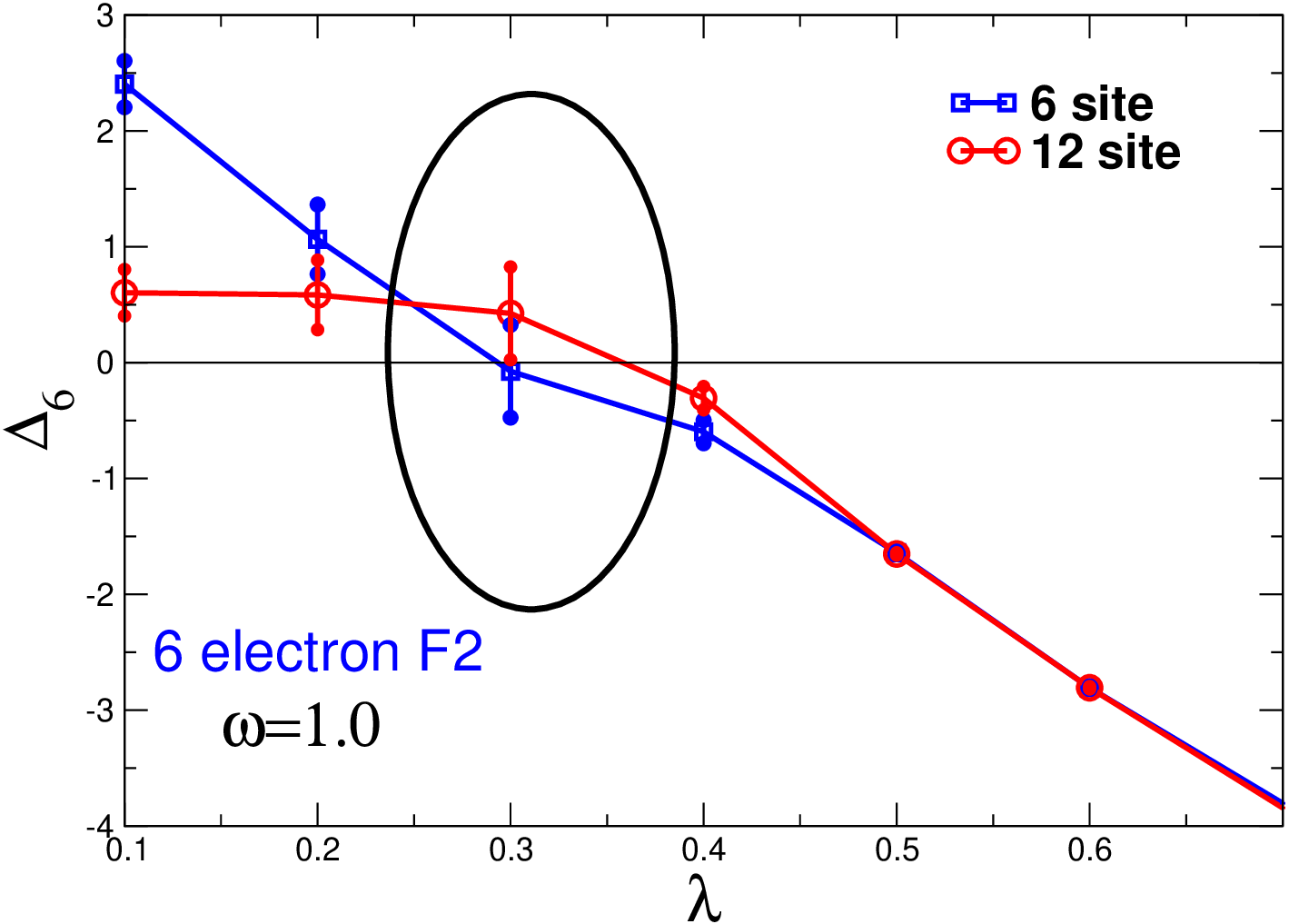} 
\caption{\label{f4} (Color online)
Binding energies for six-electron F2 lattices 
($\Delta_6$'s)
as a function of $\lambda$ ($\omega$=$1.0$ and $U$=$0$ are fixed).
Two system sizes ($L$=$6$ and $L$=$12$) are considered. 
The numerical precision is rather low in the region marked by the ellipse.
Energies are in units of $t$.
} 
\end{figure} 
%-------------------------------------------------------------------
%-------------------------------------------------------------------
%\begin{figure}[h]
\begin{figure}[b]
%\vskip 0.5cm
\includegraphics[scale=0.30]{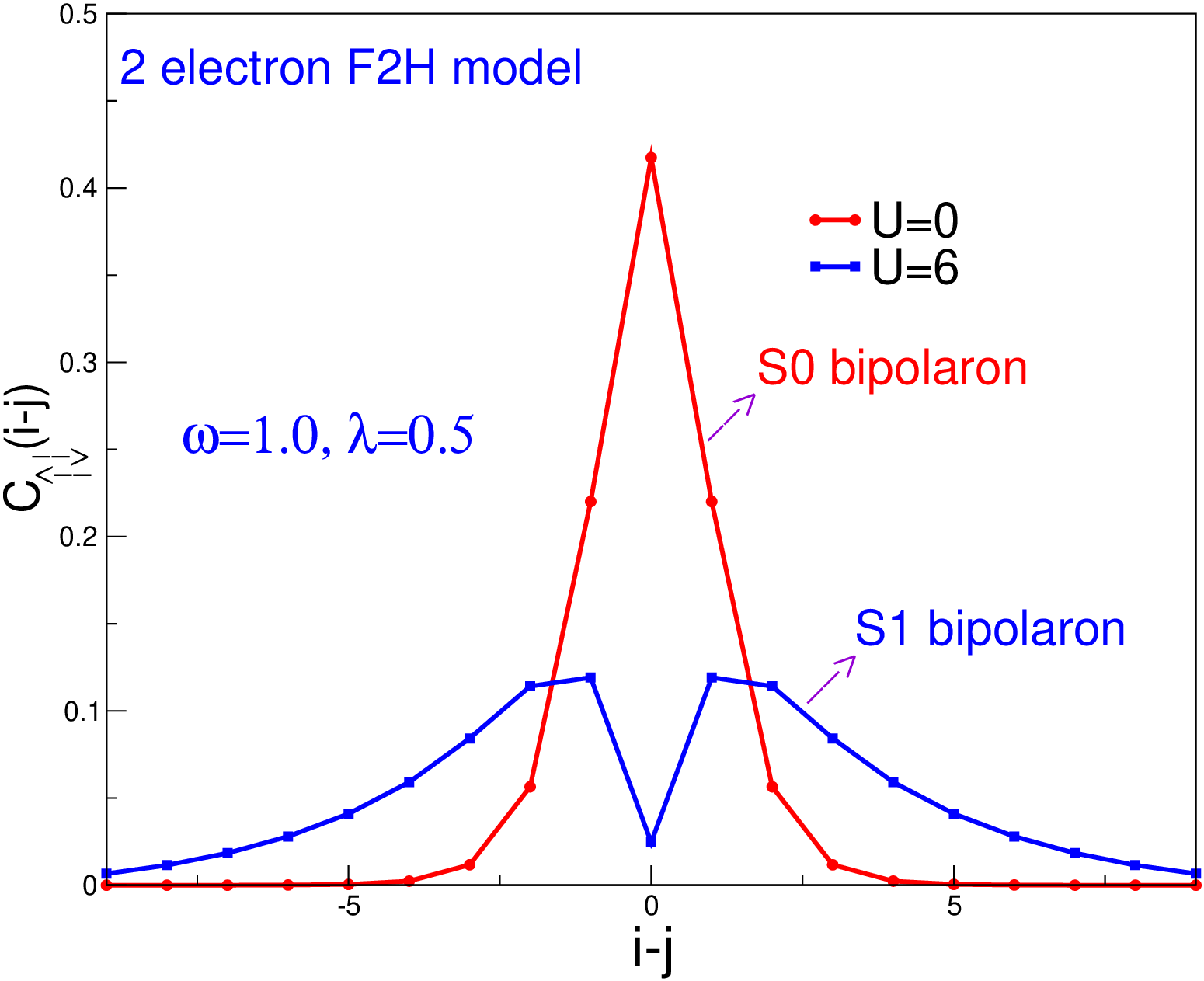} 
\caption{\label{f5}  (Color online)
The {\it e-e} correlation function $C_{\uparrow,\downarrow}(i-j)$ for
two-electron  F2H lattice. 
Two values of Hubbard $U$ are considered at $\omega$=$1.0$ and $\lambda$=$0.5$. 
Red line with circles represents the formation of an on-site $S0$ bipolaron for $U$=$0$,
while blue line with squares represents the formation of an intersite $S1$ bipolaron for $U$=$6.0$. 
The {\it b-composite} formed of two $S0$ bipolarons would have two  spin-parallel bonds,
while that formed of two $S1$ bipolarons would have either only one $S1$ spin-parallel bond or only one $S1$ spin-antiparallel bond.
} 
\end{figure} 
%-------------------------------------------------------------------

The on-site Hubbard {\it e-e} interaction $U$ has an
interesting consequence in the formation of {\it b-composite}. 
Due to the on-site nature of $U$, it is tempting to expect that $U$ does not
affect the  spin-parallel wave function that is responsible for the {\it b-b} binding.
However physics is not so simple.
Let us recall the case of  spin-antiparallel bipolaron formation for the two-electron case
of F2H model.
As shown in Fig. 5, with increasing $U$, the on-site $S0$ bipolaron formed at $U$=$0$ transforms
into intersite $S1$ bipolaron.
Now consider two cases of {\it b-b} bindings between $S0$ bipolarons and between $S1$ bipolarons.
Two $S0$ bipolarons, once they are bound, would produce 
 two  spin-antiparallel $S1$ bonds, which stabilizes the two $S1$  spin-parallel bond.
On the other hand,  above a finite $U$, the two resulting neighboring $S1$ bipolarons would have either only
one $S1$  spin-antiparallel bond or only  one $S1$  spin-parallel bond, which will not be sufficient to stabilize the {\it b-b} binding.
Therefore, with increasing $U$, the individual bipolaron transforms
from $S0$ to $S1$, which would result in the weakening of 
{\it b-b} binding between the bipolarons. 
Therefore, 
on-site $U$ interaction really influences the {\it b-b} binding, \textit{i.e.},  
with increasing $U$, the {\it b-composite} already formed above a certain $\lambda_c^{b-b}$ would dissociate 
into individual $S1$ bipolarons, and these individual bipolarons would break up into individual polarons 
with a further increase in $U$.\cite{Trug2,Mono2} 

Figure 6 shows the ground state energies of
 F2H polaron, bipolaron, and {\it b-composite} of four-electron systems as a function of $U$.
We have discussed above that, for $\lambda > \lambda_c^b$=$0.76$, an individual  spin-parallel bipolaron can be formed. 
For $\lambda$=$0.65$ ($< \lambda_c^b$) in Fig. 6(a), the energy of four-electron {\it b-composite} is the lowest at small $U$.
However, with increasing $U$, it becomes higher than twice the energy of bipolaron and then even higher than 
four times the energy of  polaron.
In contrast, for $\lambda$=$0.85$ ($> \lambda_c^b$) in Fig. 6(b), it is observed that the four-electron {\it b-composite} 
is always lower in energy than individual bipolarons and polarons.
This feature indicates that the {\it b-composite} formed above $\lambda_{c}^b$ remains glued 
even for very large $U$.

Figure 7 displays the phase diagram of F2H
polarons with respect to $\lambda$ and $U$.  There are three distinct regimes, 
(i) individual polarons, (ii) individual bipolarons, and (iii) {\it b-composite}. 
Independent bipolarons at very small $\lambda$ break into repulsive polarons
with increasing $U$.
At larger $\lambda$  and $U$=$0$, repulsive bipolarons coalesce into the {\it b-composite}.
With increasing $U$, this again dissociates into repulsive bipolaron, and then into repulsive polarons with  a further increase in $U$. 
 However, beyond a certain critical $\lambda_{c}^b=0.76$, the {\it b-composite} survives  the effect of infinite $U$.
Higher $U$ results in lighter  {\it b-composite},
 as $U$ would significantly
bring down the effective mass of the system and make it more mobile.

While the  dashed line in Fig.7 is  qualitative in nature 
because its numerical accuracy
is not very high, it should be 
noted that the phase boundary between
the polarons and bipolarons (represented by red solid line) is quite accurate 
as it is obtained from the bipolaron (2-electron) calculation.\cite{Trug2,Mono2} 
Also the vertical line representing the critical $\lambda$, above which the
stability
of {\it b-composite} is unaffected by $U$, is an analytical result by
 Bon\v{c}a and Trugman.\cite{Trug2}
 The change of curvature of the dashed line is more of numerical artifact.

%-------------------------------------------------------------------
\begin{figure}[t]
\vskip 0.5cm
\includegraphics[scale=0.30]{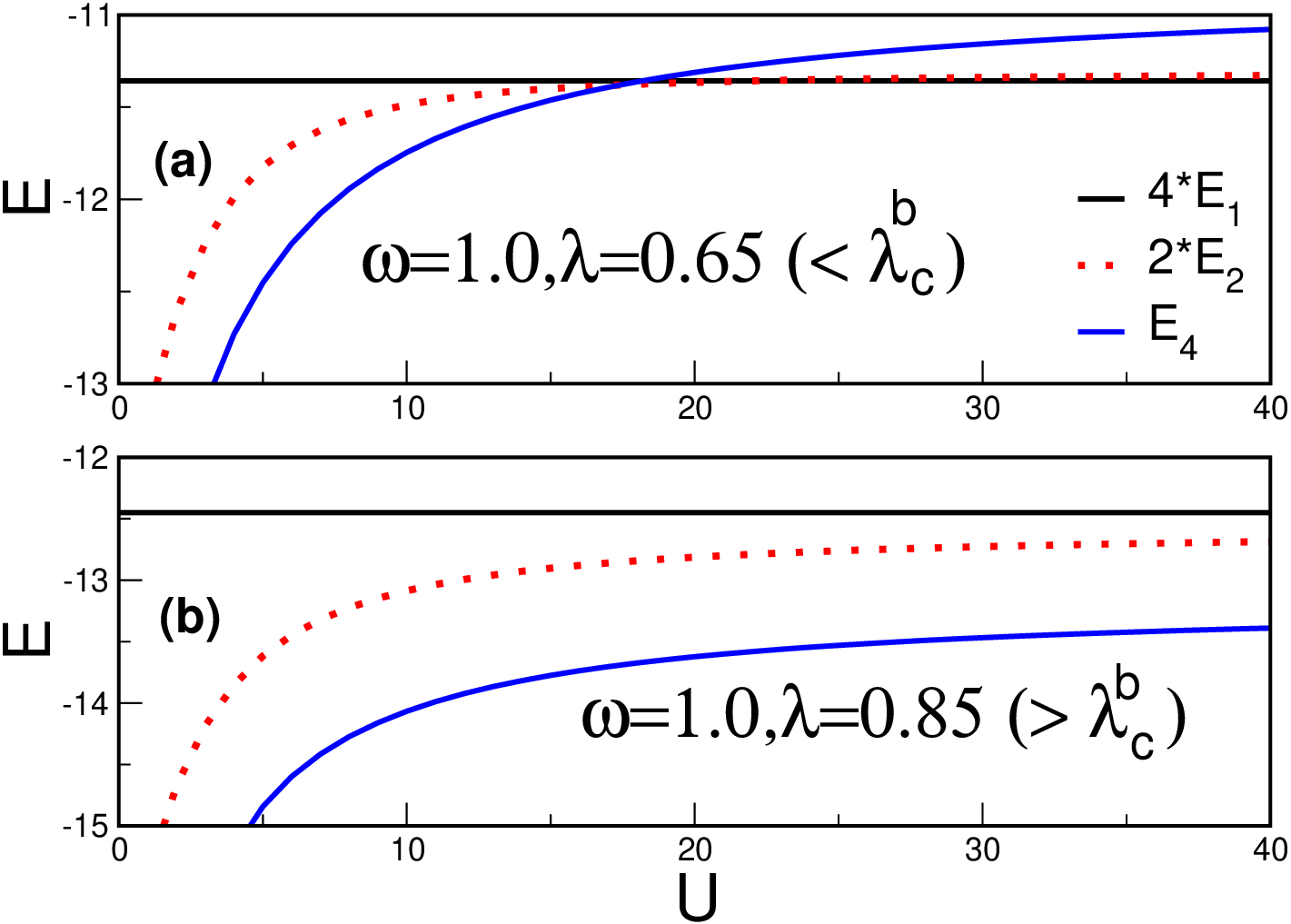} 
\caption{\label{f6}   (Color online)
Ground state energies of {\it b-composite} of four-electron ($E_4$), bipolaron ($E_2$), and polaron ($E_1$)
in the  F2H lattice ($L$=$6$) as a function of $U$. 
Blue solid line represents the
energy of {\it b-composite}, red dotted line represents twice the energy of bipolaron,
and the flat black
line represents four times the energy of polaron.
(a) Energies at $\lambda$=0.65, which is less than $\lambda_c^b=0.76$. 
(b) Energies at $\lambda$=0.85 ($> \lambda_c^b)$. 
Energy parameters are in unit of $t$, and $\omega$=$1.0$ is fixed.
} 
\end{figure} 
%-------------------------------------------------------------------
%-------------------------------------------------------------------
\begin{figure}[t]
%\vskip 0.5cm
\includegraphics[scale=0.30]{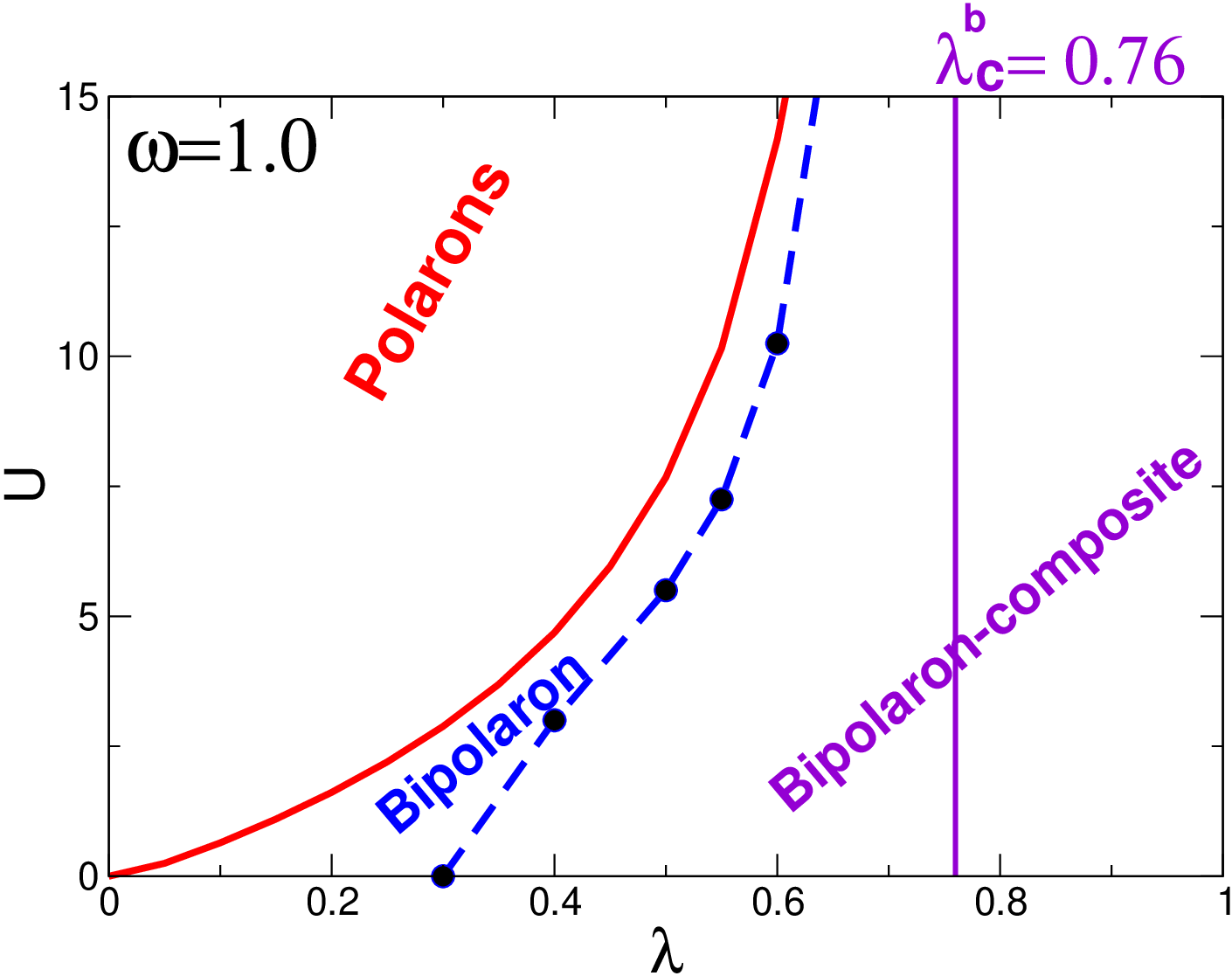} 
\caption{\label{f7}  (Color online)
Phase diagram of  F2H polarons
with respect to {\it e-ph} coupling strength $\lambda$ and on-site {\it e-e} interaction $U$ ($\omega=1.0$ is 
fixed). The blue dashed line separating the {\it b-composite} from bipolarons is a qualitative  guide for the 
eye.
}
\end{figure} 
%-------------------------------------------------------------------

          We conjecture that the formation of {\it b-composite} and its evolution with {\it U} 
are closely related to phenomena of phase separation observed in various systems,
such as CMR manganites and high $T_\mathrm{c}$ superconductors\cite{Shenoy}.
Bon\v{c}a and Trugman \cite{Trug2} suggested that, for a system with $\lambda > \lambda_{c}^b$, 
a third electron (polaron) could stick to already formed bipolaron, which would bring about the phase separation. 
Recently, Hohenadler {\it et al.}\cite{hohenalder2} also showed that the extended {\it e-ph} interaction 
yields the phase separation.
Our study corroborates this scenario.
We have demonstrated through calculations for four-electron and six-electron F2H systems
that the longer range {\it e-ph} interaction induces the glueing of bipolarons so as to produce a stable {\it b-composite},
which is reminiscent of phase separation phenomenon.
Our study thus sheds light on the microscopic description of phase separation in the presence of both {\it e-ph} and {\it e-e} interactions.

\section{Conclusions}

        We have investigated four and six-electron Holstein-Hubbard and F2H systems,
treating the electron-electron and electron-phonon interactions on an equal footing in the framework of the SC-VAED method. 
The Holstein type of on-site {\it e-ph} interaction does not support the formation of {\it b-composite},
as confirmed by both the SC-VAED and the DMRG calculations.
In contrast, the F2H type of {\it e-ph} interaction leads to the formation of {\it b-composite}
above a certain {\it e-ph} coupling strength $\lambda_c^{b-b}\approx0.3$ (at $\omega$=$1.0$).
We have shown that, with increasing Hubbard $U$, this {\it b-composite} is dissociated into individual bipolarons and then into individual polarons. 
However, above a critical $\lambda_{c}^b=0.76$ (at $\omega$=$1.0$), the {\it b-composite} survives the effect of $U$,
forming coalesced $S1$ bipolarons. 
Our unbiased study of {\it e-ph} and {\it e-e} interacting systems would provide an insight into
the understanding of phase separation physics, which is central to many areas of condensed matter physics.

%\subsection{Acknowledgement}
\acknowledgements
This work was supported by the NRF (No.2009- 0079947) and the POSTECH Physics BK21 fund. 
The work of M.T. was partially supported by the Grant-in-Aid for the
Global COE Program ``The Next Generation of Physics,
Spun from Universality and Emergence'' from MEXT of Japan. 
Stimulating discussions with Beom Hyun Kim, Bongjae Kim, and Holger Fehske are greatly appreciated.

\end{document}